\documentstyle[preprint,aps,eqsecnum]{revtex}
\def\beq{\begin{equation}}
\def\eeq{\end{equation}}

\newcommand{\bea}{\begin{eqnarray}}
\newcommand{\eea}{\end{eqnarray}}

\begin{document}
{
\tighten

\title{The New $\sigma_{tot}(\Sigma p)$ Data, the new PDG fit to hadron total 
cross sections and the TCP alternative}
 
\author{Harry J. Lipkin\,$^{a,b,c}$}
 
\address{ \vbox{\vskip 0.truecm}
  $^a\;$Department of Particle Physics \\
  Weizmann Institute of Science, Rehovot 76100, Israel \\
\vbox{\vskip 0.truecm}
$^b\;$School of Physics and Astronomy \\
Raymond and Beverly Sackler Faculty of Exact Sciences \\
Tel Aviv University, Tel Aviv, Israel \\
\vbox{\vskip 0.truecm}
$^c\;$Department of Physics, U46 \\
  University of Connecticut, 2152 Hillside Rd., Storrs, CT 06269-3046} 
 
\maketitle
 
\begin{abstract} 

The new SELEX measurement $\sigma_{tot}(\Sigma p) = 36.96 \pm 0.65$ at P =  609
GeV/c and the new 1998 Particle-Data-Group Regge (PDG) analysis of hadron total
cross sections with an additional even-signature-exchange contribution recall
the 1975 two-component-Pomeron model (TCP), which introduced such an additional
term and predicted $\sigma_{tot}(\Sigma p) = 37.07$ mb. in 1975 as well as
fitting all the same data now fit by PDG with fewer free parameters.and
predicting $\sigma_{tot}(\Sigma p)$, (not predicted by PDG) at lower energies.
The additional contribution confuses the extraction of the Pomeron intercept
from data in the 600 GeV range and its dynamical origin is still unclear. But 
its surprising systematics suggests an interesting origin.  
\end{abstract}

} 

\section{Implications of a Third Component}

The new SELEX\cite{SELEX} result $\sigma_{tot}(\Sigma p) = 36.96 \pm
0.65$ at P =  609 GeV/c, is in surprising agreement with the 1975 prediction 
$\sigma_{tot}(\Sigma p) = 37.07$ mb. from the  Two-Component-Pomeron model 
(TCP). This model arose from an analysis of the systematics of hadron-nucleon 
total cross section data\cite{NewSys} which showed the necessity of including a
new third term in addition to the commonly used Pomeron and leading Reggeon
contribuitions.  
The new accepted Particle-Data-Group (PDG) Regge analysis of hadron 
total cross sections\cite{PDG} has now also shown that three terms are needed to
fit the existing data. It is thus of interest to 
recall the TCP model which not only fits the same data with different and fewer 
parameters determined in 1975 and not changed since; it also successfully 
predicted hyperon-nucleon cross sections not predicted by PDG, now including the
new SELEX result. 

Both PDG and TCP use a Regge term which decreases with energy roughly like 
$s^{-0.5}$, and a universal Pomeron term which increases with energy roughly
like $s^{0.1}$. They also use an additional even-signature term. with an 
intermediate energy variation ($s^{-0.34}$ in the PDG model and $s^{-0.2}$
in the TCP model). Both analyses express the total hadronic cross section for 
hadron A on a proton in the form

\beq 
\sigma{Ap} = X_{Ap} s^\epsilon + Y_{1Ap}  s^{-\eta_1} 
+ Y_{2Ap} s^{-\eta_2} 
\label{sigAp}
\eeq
 PDG sets 
\beq
X^{PDG}_{\bar Ap} = X_{Ap}; ~ ~ ~ Y^{PDG}_{1\bar Ap} = Y_{1Ap}; ~ ~ ~ 
Y^{PDG}_{2\bar Ap} = - Y_{2Ap}
\label{PDGset}
\eeq
where $X_{AB}$, $Y_{1AB}$, $Y_{2AB}$, $\epsilon$, $\eta_1$, $\eta_2$
are  determined by fitting data.

 TCP  sets 
\bea
X^{TCP}_{Ap} = X\cdot N_q(A);  ~ ~ ~ Y^{TCP}_{1Ap} 
= Y_1\cdot N_q(A)\cdot N_n(A)  \nonumber \\
Y^{TCP}_{2Ap} = 
Y_2\cdot [2N_{\bar u}(A) 
 + N_{\bar d}(A)] ; ~ ~ ~ 
\eta_2 = - 0.5       
\label{TCPset}
\eea
where the coefficients $X$, $Y_1$ and  $Y_2$ are universal for all
hadrons, 
$N_q(A)$ is the total number of valence $q$ and $\bar q$  in $A$, 
$N_n(A)$ is the total number of nonstrange valence $q$ and $\bar q$  in $A$,  
$N_{\bar u}(A)$ and $N_{\bar d}(A)$ are respectively the numbers of valence 
$\bar u$ and $\bar  d$ in $A$. 

Both PDG and TCP fix parameters by fitting data, but there are  
many fewer free parameters in TCP than in PDG.
In PDG the coefficients  $X_{Ap}$, $Y_{1Ap}$ and $Y_{2Ap}$ are determined by 
fitting data and are independent of one another, except for the equality of the
isoscalar pomeron $X_{Ap}$ couplings between all states in the same isospin 
multiplet.

The particle-antiparticle relations in $Y_2$ are very different. 
In PDG  
$Y_2$ has only the odd signature $\rho$ and $\omega$ trajectories, and no 
contributions from the even signature $f$ and $A2$ trajectories.
TCP uses the known exchange degeneracy of the 
$\rho$, $\omega$, $f$ and $A2$ trajectories and therefore follows the 
Harari-Rosner\cite{HarDD} duality description in which 
 $Y_2= 0$ in exotic channels which have no resonances.

In the original TCP notation
\bea
\sigma^{TCP}_{Ap} = {{N_q(A)}\over{2}}\cdot \sigma_1 (P_{lab}/20)^\epsilon + 
\nonumber \\
{{N_q(A)\cdot N_n(A)}\over{2}}\cdot \sigma_2 (P_{lab}/20)^{-\delta} +
\nonumber \\
 + [2N_{\bar u}(A) + N_{\bar d}(A)] \cdot \sigma_R (P_{lab}/20)^{-0.5}
\label{TCPorig}
\eea
where the values determined by fitting the data available in 1975 and not 
changed since were $\sigma_1$ = 13 mb, $\epsilon$ = 0.13,  $\sigma_2$ = 4.4 mb, 
$\delta$ = 0.2 and $\sigma_R$ = 1.75 mb.  

In both PDG and TCP the exponents $\epsilon$ and $\eta_1$ are determined by 
fitting data with no theoretical input beyond the relations between different
hadrons already expressed in the formulas; i.e. the particle-antiparticle
relations in PDG and the quark-counting relations in TCP. One sees immediately
that there are many fewer free parameters in TCP and that the 
particle-antiparticle relations in the Regge term proportional to $Y_2$ are
very different between the two formulas. 

That two models with different parametrizations can fit the same data comes as 
no surprise. The data for $\sigma_{tot}(pp)$ vs. log(s) are 
very well fit by a parabola which is uniquely determined by three 
parameters\cite{PAQMREV}. Thus these data have been shown to be easily fit 
equally well by different two-Reggeon models which have four free parameters, 
two magnitudes and two exponents. 

Because the TCP couplings are universal, the expression (\ref{TCPorig}) predicts
the hyperon-nucleon cross sections with the parameters above determined in 1975
by the other experimental cross sections and no further input. 

\section{Where is the physics? What can we learn?}

     In 1975 this question was investigated by making the most naive assumptions
about the two leading terms, the Pomeron and the leading trajectories,
subtracting these contributions from the total cross sections and looking at
what remained. The surprising result, shown on fig. 4 of ref.\cite{NewSys}, is
still impressive. The additional contribution is universal above 20 GeV/c. The
$pp$, $\bar p p$, $\pi^\pm p$ and $K^\pm p$ cross sections lie on a universal
curve with scaling factors of 9:4:2 for protons, pions and kaons. This is just
the product of the total number of quarks and the number of nonstrange quarks,
the scaling factor one would obtain for a Pomeron-f cut or for a triple-regge
term in which the beam hadron couples to a Pomeron and an f. 

      What is this additional contribution? There is still no satisfactory 
explanation. But it continues to fit data and has predictive power. Note in 
particular the predictions for hyperon-nucleon cross sections then not 
available. The predicted scaling factors for $\Sigma p$ and $\Xi p$ are 6 and 3 
and they work, including the new SELEX\cite{SELEX} measurement of
$\sigma_{tot}(\Sigma p)$.

The initial motivation leading to TCP was to search  beyond the simple pole
approximation in Regge phenomenology. This first order
approximation in strong interactions could not be the whole story. The
total cross section data were already sufficiently precise to suggest a search
for new higher-order physics. The ansatz of a
double-exchange contribution to hadron-nucleon scattering with the flavor
dependence of a pomeron-f cut or a triple-Regge diagram with a pomeron and an f
coupled to the incident hadron led to a series of relations in remarkable
agreement with experiment\cite{HJLNuc}. The present situation only reinforces
the initial reaction to these results\cite{HHcom} ``I don't believe a word of
this crazy model, but the numbers are impressive. You must find a better
explanation". Since then more and more impressive numbers have been
found,\cite{NewSys,TwoPom,TwoComp,SigHyp}  but no better explanation. A 
contribution with the flavor dependence of a Pomeron-f cut and an $s$
dependence fit by a unique decreasing power fits more and more data, but there
is yet no credible explanation for this $s$ dependence. 

      The most naive assumptions used for the leading terms were that the 
Pomeron simply counts quarks and is fitted by a rising power of $s$ and that 
the leading Regge contribution counts Harari-Rosner Duality 
Diagrams\cite{HarDD}  and decrases like $s^{-(1/2)}$. Plugging these 
assumptions and fixing the five universal parameters by fitting the 1975 data 
up to 200 GeV/c gave the TCP model with the same parameters that 
still fit data accumulated since 1975. 
   
   We now examine these assumptions from the point of view of QCD.

This model can be described in modern QCD language\cite{PAQMREV} in terms of
a hierarchy of contributions inspired by large $N_c$ QCD: (1) multigluon
exchange, (2) planar quark diagrams, (3) nonplanar quark-exchange
diagrams. 

The Pomeron is described by multigluon 
exchanges which do not know
about flavor and are the same for pion and kaons and for protons and hyperons.
The additive quark counting giving the 3/2 factor between baryons and mesons is
obtained from color algebra for two-gluon and three-gluon
exchanges\cite{HJLcolor}. There is no firm justification for neglecting higher
exchanges but it fits the data. 

The leading
Regge exchanges are described Harari-Rosner duality 
diagrams are just the planar quark-exchange diagrams 
which are the leading contributions in large-$N_c$ QCD\cite{PAQMREV}.
This immediately incorporates $s-t$ duality\cite{Venez}, since exotic channels 
which have
no resonances have no contribution from planar quark diagrams.

       The third term then comes from more complicated non-planar quark
diagrams. Why these should scale in the way that they do is still open. But
this term should be absent in processes like $\phi-n$ which cannot have such
quark-exchange diagrams because there are no valence quarks in the beam and
target with the same flavor. There are no extensive data for $\phi-n$. But we
can consider as ``gedanken" $\sigma_{tot}(\phi^- p)$ the linear combination
\bea 
\sigma_{ged}(\phi^- p) \equiv
\sigma_{tot}(K^+ p)+\sigma_{tot}(K^- p)-\sigma_{tot}(\pi^- p)
\eea
which is equal
to $\sigma_{tot}(\phi^- p)$ in the quark model. The data for ``gedanken"
$\sigma_{tot}(\phi^- p)$ are shown on fig. 4 of ref.\cite{NewSys} and seen to
rise monotonically and can be fit by a single power of s as expected for a
cross section which has neither planar nor nonplanar quark exchange diagrams
and has only a Pomeron contribution. The contribution of the third term to
``gedanken" $\sigma_{tot}(\phi^- p)$ is shown on fig. 4 of ref.\cite{NewSys} to
be consistent with zero above 10 GeV/c. But the ansatz still has no convincing
basis and no firm connection with QCD beyond hand waving. 

TCP pinpoints open questions and puzzles not fully understood about the relation
between meson and baryon structure, the link between Regge phenomenology and
QCD, and how the remarkable successes of the constituent quark model can be
eventually described by QCD. 

TCP assumes $s-t$ duality\cite{Venez} in which $\sigma_{tot}(pp)$ is exotic and
has no leading Regge contribution. The decrease in $\sigma_{tot}(pp)$ with
energy observed at low energies thus indicates the existence of another
decreasing contribution in addition to leading Regge. Assuming that this
contribution is described by the double exchange ansatz and determining its
parameters by fitting $\sigma_{tot}(pp)$ then gives unique nontrivial
predictions for all other exotic cross sections; e.g. $\sigma_{tot}(K^+ p)$,
$\sigma_{tot}(\Sigma p)$, $\sigma_{tot}(\Xi p)$ and the linear combination
$\sigma_{tot}(K^- p)-\sigma_{tot}(\pi^- p)$. The linear combination ``gedanken" 
$\sigma_{tot}(\phi^- p)$  has no double exchange contribution and is predicted
to rise monotonically with the same single power of $s$ used to fit the rising
term in $\sigma_{tot}(pp)$. All predictions continue to agree with new
experimental data with no further adjustment of the five TCP parameters. 
Particularly impressive was the factor 2/3 predicted before the hyperon-nucleon
cross sections were measured which contradicted all the conventional wisdom. 

\bea
\sigma_{tot}(\pi^- p)-\sigma_{tot}(K^- p)=
(2/3)\{\sigma_{tot}(pp) - \sigma_{tot}(\Sigma p)\} =
\nonumber \\  
 = (2/3)\{\sigma_{tot}(\Sigma p) - \sigma_{tot}(\Xi p)\} 
\label{QQ1.9a}
\eea 

The remarkable success 
of naive TCP for all hadron-nucleon cross sections was summarized in
the 1981 Moriond report of the CERN hyperon experiment \cite{Exter}.

The simple systematics like the factor 3/2 between hyperon-nucleon and 
meson-nucleon strangeness differences and the monotonic rise with $s$ of the
linear combinations which have no simple quark exchange diagrams suggest the
the existence of some simple explanation based on QCD, even if the TCP ansatz
is wrong. Further investigations may provide new insight into how
QCD makes hadrons from quarks and gluons and should be encouraged.

\section{Experimental evidence that mesons and baryons are made of the 
same quarks}

The large number of relations between meson-nucleon and baryon-nucleon total
cross sections which agree with experiment suggest that mesons and baryons
are made of the same contituent quarks in the $q \bar q$ and $3 q$ 
configurations. We summarize these here and examine them from different points
of view to hopefully provide clues for theoretical explanations.

There is first the simple additive quark prediction,

\beq
 \delta_{AQM} \equiv
(2/3)\cdot \sigma_{tot}(pp) - \sigma_{tot}(\pi^-p) \leq 7\%
\eeq

There is then the TCP prediction 
\bea
\sigma_{tot}(\pi^- p)-\sigma_{tot}(K^- p) = 
\nonumber \\
 = (1/3)\sigma_{tot}(pp) - (1/2)\sigma_{tot}(K^+ p)
\eea 

Both of these are confirmed by data up to $P_{lab} = 310$ GeV/c. There are as
yet no data available for a complete set of all these reactions 
at the same single energy above $P_{lab} = 310$ GeV/c.

There are the TCP predictions for baryon-nucleon cross sections 
from meson-baryon cross sections at 100 GeV/c where data are available.

\bea
38.5 \pm 0.04 {\rm mb.} =
\sigma_{tot}(pp)= \nonumber \\
=3 \sigma_{tot}(\pi^+ p) - (3/2)\sigma_{tot}(K^- p)
= 39.3 \pm 0.2 {\rm mb.}  
\eea

\bea
33.3 \pm 0.31 {\rm mb.} = \sigma_{tot}(\Sigma p)= 
\nonumber \\ 
= (3/2)\{\sigma_{tot}(K^+ p) +
\sigma_{tot}(\pi^- p)-\sigma_{tot}(K^- p)\}  =
\nonumber \\ 
= 33.6 \pm 0.16 {\rm mb.}  
\eea

\bea
29.2 \pm 0.29 {\rm mb.} = \sigma_{tot}(\Xi p) = 
\nonumber \\
= (3/2)\sigma_{tot}(K^+ p) 
= 28.4\pm 0.1 {\rm mb.}  
\eea

\bea
\sigma_{tot}(\Omega^- p)= 
\nonumber \\ 
=(3/2)\{\sigma_{tot}(K^+ p) -
\sigma_{tot}(\pi^- p)+\sigma_{tot}(K^- p)\}  
\eea

Another interesting way to view the data is to compare the 
strange and nonstrange quark contributions to the to baryon-nucleon and 
meson-nucleon total cross sections extracted using the additive quark model 

Let $\sigma(fN)_H$ denote the total cross section on a nucleon target, for a
single quark of flavor $f$ in a hadron $H$ on a nucleon target, where $f$ may
be strange $s$ or nonstrange $n$ and $H$ may be a baryon $B$ or a meson $M$.
The additive quark model gives 

\beq
\sigma(nN)_B = {1\over 3}\cdot \sigma(pN)
= 12.9  \pm  0.01 mb.
\eeq

\bea
\sigma(sN)_B = {1\over 3}\{\sigma(\Sigma N)+\sigma(\Xi N)-\sigma(pN)\} =
\nonumber \\ 
= 7.7 \pm 0.1 mb.
\eea

\bea
\sigma(nN)_M = {1\over 2}\{\sigma(\pi N)-\sigma(\bar K N)
+\sigma(K N)\}m =
\nonumber \\ 
= 11.2 \pm  0.05 mb.
\eea

\bea
\sigma(sN)_M = {1\over 2}\{\sigma(\bar K  N)-\sigma(\pi N)
+\sigma(K N)\} =
\nonumber \\ 
=  7.75  \pm  0.05 mb.
\eea
where we have assumed $   \sigma(sN)_M = \sigma(\bar s N)_M  $.

We find the surprising result that the contribution of the strange quarks
is the same to both meson-nucleon and baryon-nucleon total ross sections,
but that the contribution of the nonstrange quarks
is the less for meson-nucleon than for baryon-nucleon total cross sections.
This immediately gives rise to speculations that nonstange quarks are more
complicated than strange quarks because they can have a pion cloud. However. 
there has been no success in carrying this argument further quantitatively.
Instead we obtain the following surprising relations, which go to the heart of 
the TCP ansatz; namely that one single mechanism is responsible for the 
breakings of both SU(3) flavor symmetry and the additive quark model.
\beq
\sigma(nN)_B - \sigma(nN)_M =  1.69 \pm 0.05 mb.
\eeq

\beq
{1\over 2}\{ \sigma(nN)_M - \sigma(sN)_M\} = 1.73 \pm 0.04 mb.
\eeq

The difference between the contributions of nonstrange quarks to baryon-nucleon
and meson-nucleon cross sections is equal to the difference between the 
contributions of nonstrange and strange quarks quarks to meson-nucleon cross
sections.  

This as yet unexplained connection between the 
deviation from SU(3) symmetry and 
the deviation from the Levin-Frankfurt AQM 3/2 ratio for baryons and mesons
has been expressed by the experimentally satisfied relation\cite{NewSys} 
\beq
\sigma_{tot}(\pi^- p)-\sigma_{tot}(K^- p)=
(1/3)\sigma_{tot}(pp) - (1/2)\sigma_{tot}(K^+ p)
\eeq
This has been rearranged to give 
\bea
\sigma_{ged}(\phi^- p) \equiv \sigma_{tot}(K^+ p) + \sigma_{tot}(K^- p)
- \sigma_{tot}(\pi^- p)= 
\nonumber \\ 
= (3/2)\sigma_{tot}(K^+ p) - (1/3)\sigma_{tot}(pp) 
\eea
The expressions on both sides of this relations are found experimentally
not  only to be equal but to increase monotonically with energy and fit by
a single power. This fits in with the picture that 
$\sigma_{ged}(\phi^- p)$ contains only a pure Pomoeron contribution. But
that the right hand side which is a linear combination of meson and baryon
cross sections behaves in the same way suggests some sort of universality for
the Pomeron.
  
We also find that
the difference between the contributions of nonstrange and strange quarks 
to baryon-nucleon cross sections is greater by a factor of (3/2) than the 
corresponding difference for meson-nucleon cross sections.
\beq
\sigma(nN)_B - \sigma(sN)_B = 5.15 \pm 0.07 mb. 
\eeq

\beq
{3\over 2}\{ \sigma(nN)_M - \sigma(sN)_M\} = 5.2 \pm 0.1 mb.
\eeq

As soon as one begins to think about some dynamical origin for these relations
one encounters a very perplexing question. Why is there seemingly such a simple
relation between hadron-nucleon total cross sections, when any credible
scattering model suggests that they should be very different and depend upon
radii and geometrical considerations, not simply quark counting? Perhaps a
relevant interesting analogy is in the relation between hadronic elecromagnetic
form factors and eloectric charges. The form factors of pions, nucleons and
other hadrons are complicated and very different from one another. But their
total electric charge is simple and given by adding up the charges of their
constituent quarks. Rutherford scattering measures total charge. Does
universality of contributions to $\sigma_{tot}(H p) $ suggest measurements of
some kinds of total charge in which the microscopic details are smehow not
important? 

\section{Conclusions} 

We conclude by listing a hierarchy of the experimental systematics found in
this phenomenological analysis and the questions to be resolved by further 
experiments at higher energies:

\subsection {Summary of experimental systematic regularities}

\noindent 1. Odd signature universality

   $\rho$ universality (Sakurai) - conserved isospin current

   $\omega$ universality - related to $\rho$ by U(2)
 
   Energy dependence - like $s^{-1/2}$

\noindent 2. Exchange degeneracy - No exotic contributions
   
   Only planar quark diagrams contribute

\noindent 3. Universal Pomeron - Counts quarks

   Given by amplitude with no quark exchanges

   $\sigma_{ged}(\phi^- p) = 
\sigma_{tot}(K^+ p)+\sigma_{tot}(K^-p)-\sigma_{tot}(\pi^- p) $ 

\noindent 4. What's left?

Universal contribution scaling like Pomeron-f cut

Scaling factors of 9:4:2 like for protons, pions and kaons. 

Extrapolated to 6 and 3 for $\Sigma p$ and $\Xi p$ - predict data!

But what is it?

\subsection {Interesting questions to be decided by future experiments}

\noindent 1. Does the ad-hoc third component continue to explain both the 
deviation from SU(3) symmetry and 
the deviation from the Levin-Frankfurt AQM 3/2 ratio for baryons and mesons; 
i.e. do 
$ \sigma_{tot}(\pi^- p)-\sigma_{tot}(K^- p)$ and 
$(1/3)\sigma_{tot}(pp) - (1/2)\sigma_{tot}(K^+ p)$
continue to remain equal at higher energies?

\noindent 2. Do both the SU(3) breaking and the deviation from 3/2 go to zero
at high energies or does one or both level off. Data at around 200 GeV/c
indicate that $ \sigma_{tot}(\pi^- p)-\sigma_{tot}(K^- p)$ might be leveling
off while $(1/3)\sigma_{tot}(pp) - (1/2)\sigma_{tot}(K^+ p)$ continues to
decrease with increasing energy.  But the differences are not convincing and
better data at higher energies should easily resolve this question. 

\noindent 3. Is there a universal Pomeron that holds for all hadrons with a 3/2
ratio between baryon and meson couplings? Do $\sigma_{ged}(\phi^- p)$
and  $(3/2)\sigma_{tot}(K^+ p) - (1/3)\sigma_{tot}(pp)$ continue to be equal
and rise monotonically like a pure Pomeron conmtribution? To answer this
question reliably one must go to high enough energies so that the contribution
of the third component becomes negligible. 

\subsection {Bottom Line}

There is much yet to learn from future experiments about how QCD makes hadrons 
out of quarks and gluons.
 
\acknowledgments
It is a pleasure to thank Edmond Berger, Uwe Dersch, Joe Lach 
and Murray Moinester for helpful discussions and comments. This work
was partially Supported by a grant from US-Israel Bi-National Science 
Foundation.
 
{
\tighten

}

\end{document}